\documentclass[11pt,oneside,a4paper]{article}
\usepackage{amsmath}
\usepackage{verbatim}
\usepackage[ansinew]{inputenc}
\usepackage[dvips]{color,graphicx}
\usepackage{subfigure}
\usepackage{setspace}

\begin{document}
\begin{center}
{\Large Correlated Counting of Single Electrons in a Nanowire Double Quantum Dot}
\end{center}
\begin{center}
Theodore Choi$^{1}$, Ivan Shorubalko$^{1}$, Simon Gustavsson$^{1}$, Silke Schön$^{2}$, and Klaus Ensslin$^{1}$
\end{center}
\begin{center}
\textit{$^{1}$Solid State Physics Laboratory, ETH Zurich, 8093 Zurich, Switzerland}
\end{center}
\begin{center}
\textit{$^{2}$FIRST lab, ETH Zurich, 8093 Zurich, Switzerland}
\end{center}
\begin{abstract}
We report on correlated real-time detection of individual electrons in an InAs nanowire double quantum dot. Two self-aligned quantum point contacts in an underlying two-dimensional electron gas material serve as highly sensitive charge detectors for the double quantum dot. Tunnel processes of individual electrons and all tunnel rates are determined by simultaneous measurements of the correlated signals of the quantum point contacts.
\end{abstract}
\tableofcontents
\section{Introduction}
Probing the electronic state of quantum dots (QDs) by charge detection with quantum point contacts (QPCs) represents an elaborate method for investigating tunnel processes of single charges \cite{field93}. The high sensitivity of the conductance of a QPC to its electrostatic environment allows to measure transitions in QDs in a regime where conventional transport measurements are impeded by the limited resolution of standard current meters. However, charge detection measurements are usually based on monitoring the average conductance of the QPC and thus measuring only the change of the average population of the QDs. In this respect, time-resolved charge detection marks a significant improvement since it offers the ability to count tunnel processes of individual single charges in real-time \cite{schleser04,vandersypen04}.

Significant achievements on the time-resolved detection of single charges have been done in metallic structures. As an example, time-resolved charge detection has been used to measure time-correlated tunneling events of single charges in a one-dimensional array of metallic tunnel junctions using a resistively coupled SET as a charge detector \cite{bylander05}. This method has allowed to demonstrate experimentally the relation between frequency and current given by $I=ef$ and has suggested the possible application for quantum metrology.

In this work, we use semiconductor nanowires (NWs) which are promising candidates for electronic nanoscale devices and for studying fundamental physics of low-dimensional systems \cite{bjoerk04,bjoerk02}. InAs is a particularly interesting material since it exhibits large confinement energies due to the small effective mass of the electrons. Thus, experimental access for probing quantum states is significantly facilitated. Furthermore, a large effective $g^{\star}$ factor enhances the ability for magnetic control of spins. At the same time, strong spin-orbit interaction known for InAs could be exploited for the manipulation of single spins by electric fields \cite{golovach06,bulaev07,flindt06}. These specific properties make InAs a unique material for realizing quantum bits in solid-state-based quantum computers by spins in coupled quantum dots \cite{loss98}.

QDs have been fabricated in InAs NWs by top gates \cite{pfund06}, etching \cite{shorubalko08}, local back gates \cite{fasth05} and by including layers of InP \cite{bjoerk04}. Charge readout has been reported on a double quantum dot (DQD) in a Ge/Si core/shell heterostructure NW by capacitively coupling the DQD to a nearby QD serving as a charge detector \cite{hu07}.

Here, we present time-resolved charge detection measurements with directional resolution \cite{fujisawa06} on a tunable etched InAs NW DQD using the correlated signal of two self-aligned QPCs serving as highly sensitive charge detectors. We determine all tunnel rates of the tunnel processes from the time traces of the QPC signals. The strong detection signal and the advantageous properties of InAs offer the prospect for time-resolved detection of individual spins \cite{elzerman04}.

\section{Experimental Setup}
The NWs are grown by metal organic vapor-phase epitaxy (MOVPE) using colloidal Au particles as catalysts \cite{pfund06b}. The NWs have wurtzite crystal structure and are typically $100~$nm in diameter and $10~\mu\textrm{m}$ long. The NWs are deposited on a predefined Hall bar of an AlGaAs/GaAs heterostructure with standard ohmic contacts. The heterostructure is grown by MBE and contains a 2DEG $37~$nm below the surface. The NW is furnished with Ti/Au ohmic contacts. EBL patterned PMMA is used as an etching mask to define the DQD and the QPCs. The structure is etched using a H$_{2}$O/H$_{2}$SO$_{4}$/H$_{2}$O$_{2}$ (100:3:1) solution with an etching rate of $\sim 1.7~$nm/s for both the NW and the heterostructure substrate. Typical etching times are around $15~$s. The structure is designed in such a way that the trenches in the 2DEG forming the QPCs by depletion of the 2DEG underneath and the constrictions in the NW forming tunnel barriers of the DQD are defined in a single step etching process. The fact that both trenches of the 2DEG and constrictions of the NW are defined simultaneously by the same etching areas ensures perfect alignment of the QPCs and the QDs.\\
\begin{figure}[h]
	\centering
		\includegraphics[width=0.65\textwidth]{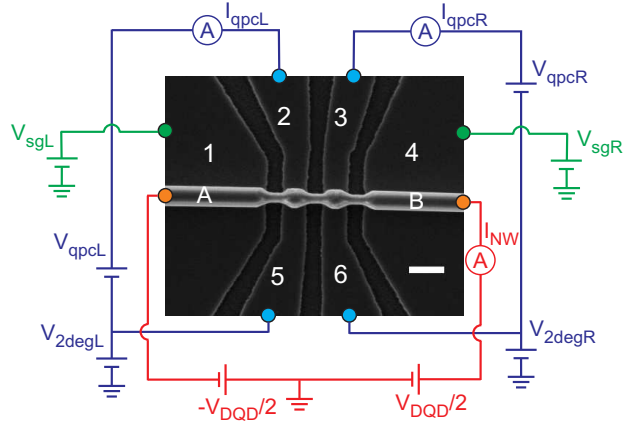}
	\caption{SEM image of the etched DQD structure with QPCs and lateral side gates. A circuit scheme with the voltages and currents used in the measurement is shown in addition. A/B are the source/drain contacts of the NW. The source/drain contacts of the QPCs are the contacts 2/5 and 3/6 for the left and right QPC respectively. The electronic width of the QPCs can be changed by applying voltages to the side gates on contacts 1/4. The current is measured through both QPCs and the NW. The bias voltage on the DQD is applied symmetrically to source and drain. Scale bar: $200~$nm.}
	\label{fig:DQD_circuit}
\end{figure}
\\

An SEM image of the etched DQD structure and self-aligned QPCs together with a circuit scheme is shown in Fig.\ref{fig:DQD_circuit}. The QPCs operate as local gates to change the electron population in each QD and as sensitive charge detectors for transitions in the DQD. Compensation voltages $V_{\textrm{sgL/R}}$ were applied to the side gates in order to keep both QPCs at a constant operation point. For the presented measurements, the QPCs are operated at a slope of the conductance close to pinch-off, where we get a desirable sensitivity to transitions in the DQD. All measurements presented here are dc measurements and were performed at $T=1.8~$K.

\section{Time-averaged charge detection}
\begin{figure}[h]
	\centering
		\includegraphics[width=1\textwidth]{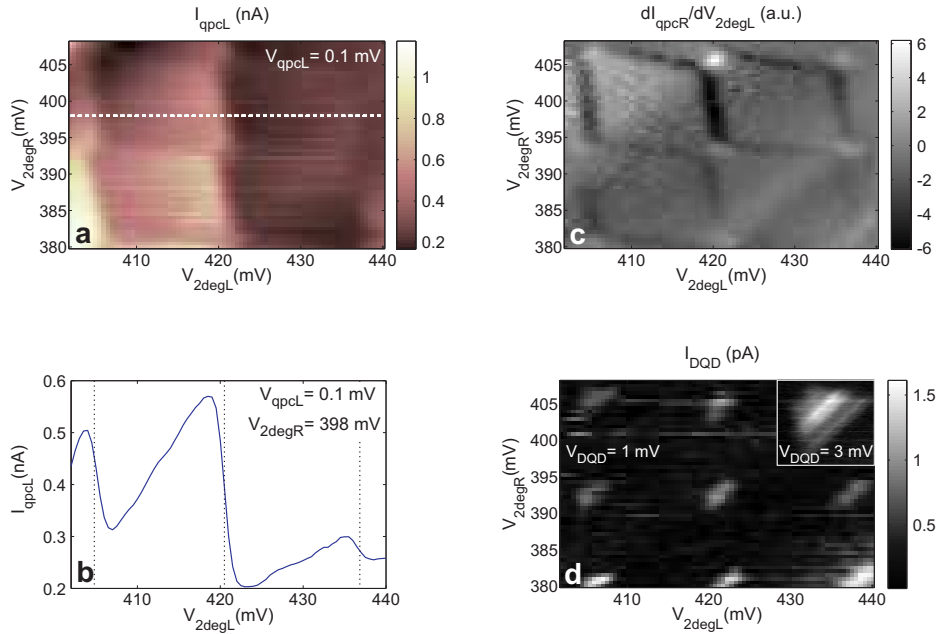}
	\caption{Simultaneous measurements of the charge stability diagram by charge readout of both QPCs and the source-drain current through the DQD. (a) Current $I_{\textrm{qpcL}}$ along the left QPC as a function of the gate voltages $V_{\textrm{2degL/R}}$. (b) Cut through the dashed line in (a). (c) Numerical transconductance $dI_{\textrm{qpcR}}/dV_{\textrm{2degL}}$ of the right QPC versus the gate voltages $V_{\textrm{2degL/R}}$. (d) Source-drain current through the DQD for the same region of gate voltages.}
	\label{fig:honeycombs_dc}
\end{figure}

Fig.\ref{fig:honeycombs_dc}(a) shows the current through the left QPC as a function of the gate voltages $V_{\textrm{2degL/R}}$. The bias voltages are $V_{\textrm{qpcL}}=0.1~$mV, $V_{\textrm{DQD}}=1~$mV. The honeycomb diagram reflecting the charge states of the DQD can clearly be recognized. A cut along the dashed line in Fig.\ref{fig:honeycombs_dc}(a) shows that the QPC current $I_{\textrm{qpcL}}$ exhibits a distinct step whenever the left QD is populated by an additional electron (Fig.\ref{fig:honeycombs_dc}(b)). The strong capacitive coupling of the QPCs to the DQD leading to a large relative change in QPC conductance of up to $66~\%$ by the addition of a single electron to the DQD demonstrates the exceptional capability for charge detection in the present system. Furthermore, the device containing two QPCs allows to probe the charge state of the DQD with both QPCs at the same time. Fig.\ref{fig:honeycombs_dc}(c) shows the transconductance $dI_{\textrm{qpcR}}/dV_{\textrm{2degL}}$ of the right QPC versus the gate voltages $V_{\textrm{2degL/R}}$. The honeycomb diagram detected by the right QPC matches exactly the measurement of the left QPC in Fig.\ref{fig:honeycombs_dc}(a), showing transitions in both QDs with comparable sensitivity. The reversed signs of the transconductance for the vertical/horizontal boundary lines of the honeycomb cells and for the transitions at the triple points reflect the QD-lead or the interdot transitions. The reliability of the charge detection method using the QPCs is illustrated by a simultaneous measurement of the source-drain current $I_{\textrm{DQD}}$ through the DQD shown in Fig.\ref{fig:honeycombs_dc}(d). Elastic tunneling through the DQD resulting in a high current $I_{\textrm{DQD}}$ is only possible at the triple points where the electrochemical potentials of the QDs align with the Fermi level in the leads. Enhanced current can clearly be seen at the triple points exactly in line with the corners of the honeycomb cells from the QPC signals in Fig.\ref{fig:honeycombs_dc}(a) and (c). 

Applying a finite bias to the DQD causes the triple points to develop into triangular shaped regions inside which the current through the DQD is given by inelastic tunnel processes (inset in Fig.\ref{fig:honeycombs_dc}(d)) \cite{vanderwiel03}. Lines of enhanced current inside the finite bias triangles representing excited states of the QDs allow to determine the level spacing of the first excited states from the QD ground states, yielding $\Delta E_{L}\approx1.4~$meV and $\Delta E_{R}\approx1.2~$meV for the left and right QD, in agreement with an estimation of the spacing assuming spherical QDs with hard walls. 

Using a capacitive model to describe the interactions between the QDs and the environment \cite{vanderwiel03}, all capacitances and lever arms as well as charging energies of the system are extracted from the dimensions of the honeycomb cells and the finite bias triangles. The total capacitances of the QDs are $C_{\Sigma L}=25.6~$aF and $C_{\Sigma R}=40.9~$aF for the left and the right QD, respectively. The mutual capacitance between the QDs is $C_{m}=3.5~$aF. The lever arms for conversion of gate voltages to energy are $\alpha_{L}=0.39$ and $\alpha_{R}=0.34$. Finally, we obtain for the charging energies of left and right QD $E_{CL}=6.3~$meV and $E_{CR}=4.0~$meV, and a mutual charging energy of $E_{Cm}=0.54~$meV. From cuts of the QPC current through the vertical/horizontal boundary lines of a honeycomb cell we determine an electron temperature of $T=1.8~$K in the leads. Extraction of the tunnel coupling $t$ from the width of the transition at the triple points \cite{Dicarlo04} is impeded by the fact that broadening due to temperature dominates in the present measurement.\\
\newpage
\section{Time-resolved charge detection}
\begin{figure}[h]
	\centering
		\includegraphics[width=1\textwidth]{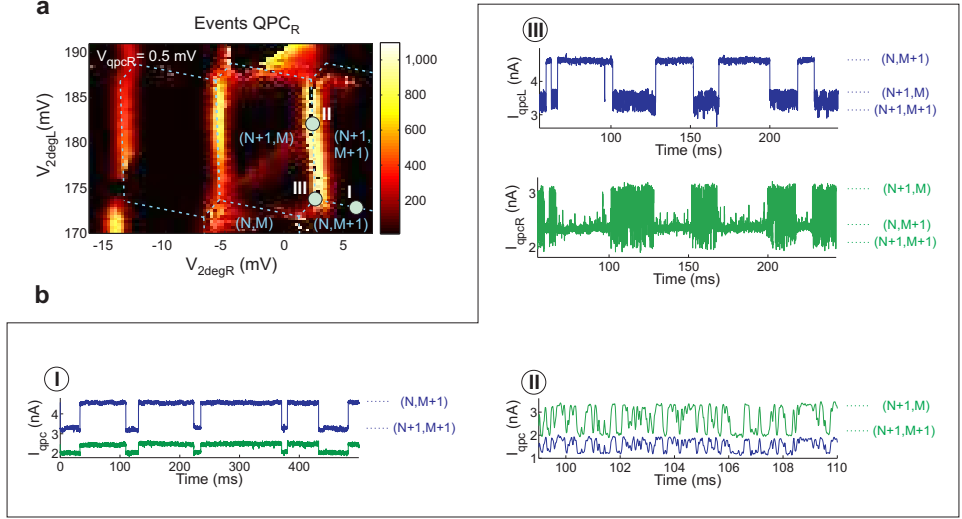}
	\caption{Measurement of the honeycomb cells by time-resolved charge detection. (a) Counting events of the right QPC as a function of the gate voltages $V_{\textrm{2degL/R}}$. (b) Time traces of both QPCs taken at selected positions in (a). Current traces of the right QPC are shown in green, the traces of the left QPC in blue.}
	\label{fig:honeycomb_counting}
\end{figure}

Time-resolved measurements of the tunneling processes in the DQD become possible as soon as the tunnel rates of the barriers defining the DQD are below the bandwidth of $20~$kHz of the measurement setup \cite{gustavsson08}. We tune the tunnel rates of the barriers in the NW by moving to different gate voltage regions until the tunnel processes occur on a sufficiently slow time scale. Single electrons tunneling through the three barriers defining the DQD can then be counted one by one in real-time. In contrast to previous experiments \cite{fujisawa06} we measure the two QPC signals simultaneously whose (anti-) correlation gives additional information on the electron tunneling processes. Fig.\ref{fig:honeycomb_counting}(a) shows a honeycomb cell measurement where the tunnel events are detected by recording time traces of the right QPC current at each point of the swept range of gate voltages. A schematic drawing of the honeycomb cells is given as a guide for the eye. We denote the charge state of the DQD by $(N,M)$, where $N$ and $M$ are the number of electrons on the left and the right QD respectively. The length of the time traces is $0.5~$s. The bias on the QPC is $0.5~$mV while no bias is applied on the DQD. A high number of events is detected along the boundary line of the honeycombs where the electrochemical potential of the right QD is aligned with the Fermi level of the drain lead. Events corresponding to transitions between the left QD and the source lead through the left barrier occur on a much lower rate ($\sim 10~$events/s) and are not visible on the color scale shown in the figure. Thus, the tunnel rates of the barriers connecting the DQD to source and drain lead are strongly asymmetric obeying the relation $\Gamma_{\textrm{L}}\ll \Gamma_{\textrm{R}}$.\\
\\ 

Simultaneous to the charge detection with the right QPC, counting was also performed with the left QPC. Time traces of the current of both left and right QPC taken at selected positions in Fig.\ref{fig:honeycomb_counting}(a) allow a more comprehensive analysis of the tunnel processes across the DQD. Fig.\ref{fig:honeycomb_counting}(b)(I) shows a time trace of the QPC signals at a position where the electrochemical potential of the left QD is aligned with the source lead and tunneling thus occurs across the left barrier. The green traces are the current traces of the right QPC whereas the left QPC traces are shown in blue. Both the left and the right QPC signals can clearly resolve two current levels corresponding to the charge states $(N,M+1)$ and $(N+1,M+1)$. The time trace for the situation where the electrons tunnel back and forth across the right barrier shows also a switching between two current levels corresponding to the $(N+1,M)$ and $(N+1,M+1)$ charge states, but on a much shorter time scale (Fig.\ref{fig:honeycomb_counting}(b)(II)). The time traces representing the transitions across the two barriers connecting the DQD to the leads thus confirm the asymmetry of the tunnel rates of the barriers. In addition, we observe perfect correlation of the left and right QPC signals proving that the same transition processes are detected by both QPCs. The time traces of the QPC currents allow to extract the tunnel rates $\Gamma_{\textrm{L/R}}$ of the barriers. We obtain for the left barrier $\Gamma_{\textrm{L}}=48~$Hz and accordingly for the right barrier $\Gamma_{\textrm{R}}=15.1~$kHz.

An interesting behavior is observed for the time trace taken at a triple point shown in Fig.\ref{fig:honeycomb_counting}(b)(III). Here, the three charge states $(N,M+1)$, $(N+1,M)$ and $(N+1,M+1)$ are degenerate. Three current levels corresponding to the degenerate charge states can be distinguished in both QPC current traces. Strikingly, both QPC signals exhibit a bunching of fast switching events interrupted by transitions occuring on a much slower time scale. Furthermore, perfect correlation of the left and right QPC signals is observable. We relate the fast transitions to equilibrium fluctuations across the transparent right barrier where the charge of the right QD changes by one. Those fast transitions alternate with occasional switches of the QPC signal to the $(N,M+1)$ level. This can happen in two ways: (i) Starting from the configuration $(N+1,M)$ with one excess electron in the left QD, the electron tunnels through the central barrier to the right QD such that the DQD charge state ends up in $(N,M+1)$. (ii) The excess electron in the left QD tunnels from the $(N+1,M+1)$ charge state through the left barrier to the source lead leaving the DQD in the $(N,M+1)$ state. Both processes can be observed in the time trace at the considered triple point and we distinguish them by examining the particular course of the QPC signals at the transitions to the $(N,M+1)$ current level.

\begin{figure}[h]
	\centering
		\includegraphics[width=0.70\textwidth]{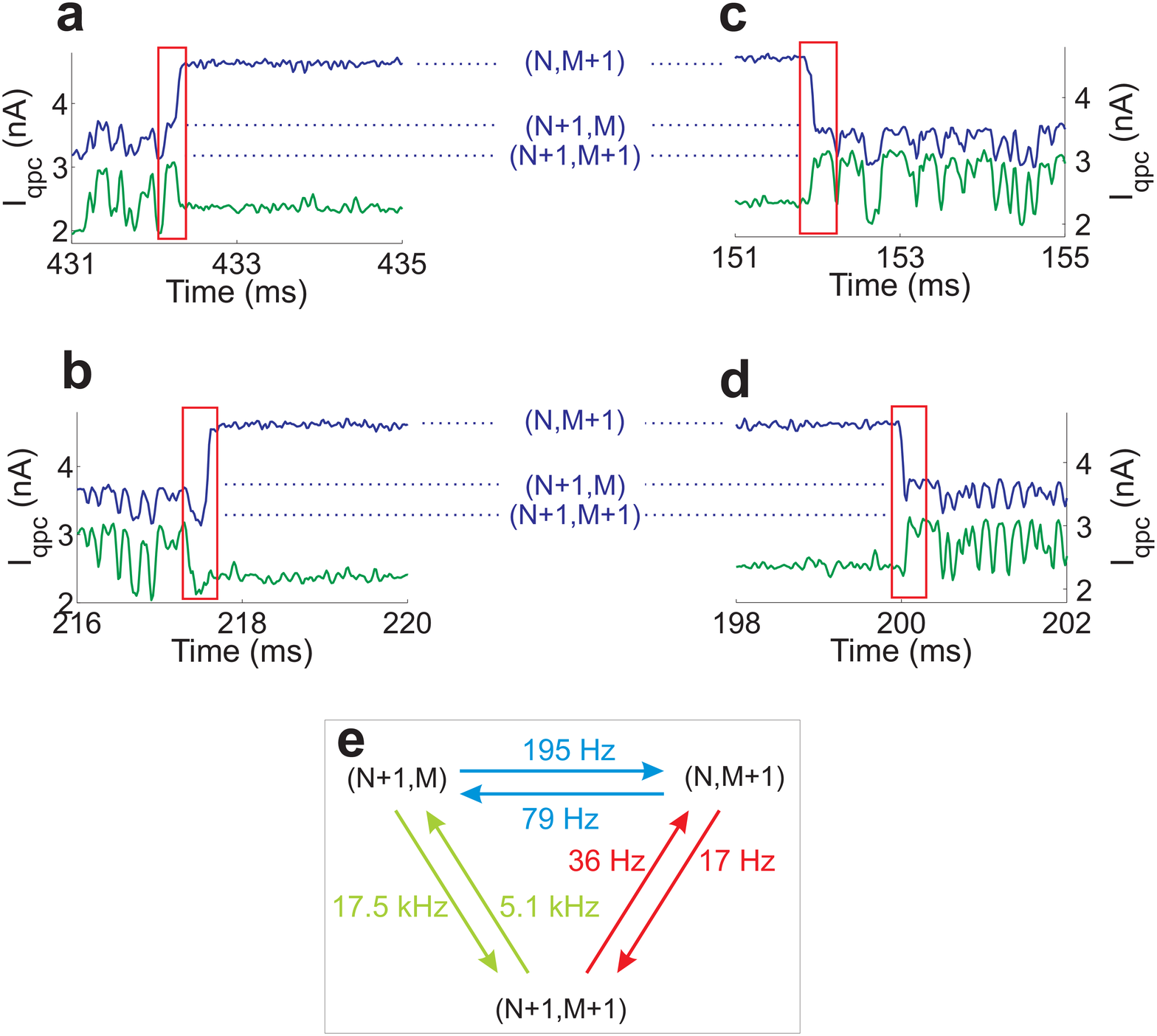}
	\caption{(a)-(d) Close-ups to transitions in the time trace of Fig.\ref{fig:honeycomb_counting}(b)(III). The red boxes frame the transitions discussed in the text. (e) Schematic representation showing all involved tunnel processes with the corresponding tunnel rates $\Gamma_{(I,J)\rightarrow(K,L)}$.}
	\label{fig:timetraces_zoom}
\end{figure}

Fig.\ref{fig:timetraces_zoom} shows four close-ups to the time trace of Fig.\ref{fig:honeycomb_counting}(b)(III) at positions where the current switches from the bunching of fast transitions to the $(N,M+1)$ state and back. We observe the process (i) in Fig.\ref{fig:timetraces_zoom}(a). The fast transitions between the $(N+1,M)$ and $(N+1,M+1)$ states are perfectly correlated for both QPC signals. While switching from the $(N+1,M+1)$ to the $(N,M+1)$ level, the left QPC signal exhibits a step at the $(N+1,M)$ level. Thus we observe the transport sequence $(N+1,M+1) \rightarrow (N+1,M) \rightarrow (N,M+1)$. The transition from the $(N+1,M+1)$ to the $(N+1,M)$ state results in a correlated switching of both QPC signals since the electron is tunneling out of the right QD to the drain lead. The subsequent interdot transition from the $(N+1,M)$ to the $(N,M+1)$ state however, gives an anticorrelated signal of the QPCs (red box in Fig.\ref{fig:timetraces_zoom}(a)). Since the electron is moving from the left to the right QD, the conductance of the left QPC will increase while the conductance of the right QPC decreases. The process (ii) is shown in Fig.\ref{fig:timetraces_zoom}(b). Here the transport sequence at the transition to the $(N,M+1)$ state is given by $(N+1,M) \rightarrow (N+1,M+1) \rightarrow (N,M+1)$. The QPC signals are correlated since no interdot transition is involved in the sequence (red box in Fig.\ref{fig:timetraces_zoom}(b)). Fig.\ref{fig:timetraces_zoom}(c) and (d) show the same transport sequences as in (a) and (b) but in reversed tunneling direction. Thus, monitoring the current of both QPCs allows to track a single electron tunneling through the DQD in real time and with directional resolution.\\
\\

The average time $\langle \tau_{(I,J)} \rangle$ that the DQD spends in the $(I,J)$ charge state is related to the tunnel rates $\Gamma_{(I,J)\rightarrow(K,L)}$ from the $(I,J)$ to the $(K,L)$ state by $\langle \tau_{(I,J)} \rangle^{-1}=\sum_{(K,L)}\Gamma_{(I,J)\rightarrow(K,L)}$, assuming an exponential distribution of the tunneling times $ \tau_{(I,J)}$ \cite{fujisawa06}. Here, we calculate all tunnel rates of transitions between the three charge states $(N,M+1)$, $(N+1,M)$ and  $(N+1,M+1)$ both by determining the average time the DQD spends in a particular charge state and by counting the number of transitions $n_{(I,J)\rightarrow(K,L)}$ from the $(I,J)$ to the $(K,L)$ state. The tunnel rate $\Gamma_{(I,J)\rightarrow(K,L)}$ can be extracted from $n_{(I,J)\rightarrow(K,L)}$ by means of the relation $n_{(I,J)\rightarrow(K,L)}=p_{(I,J)}\Gamma_{(I,J)\rightarrow(K,L)}\Delta t$, where $p_{(I,J)}$ is the occupation probability of the $(I,J)$ state and $\Delta t$ the length of the time trace. Fig.\ref{fig:timetraces_zoom}(e) shows a schematic representation of all the observed transitions with the corresponding tunnel rates $\Gamma_{(I,J)\rightarrow(K,L)}$. Thus, the tunnel rates of the barriers defining the DQD obey the relation $\Gamma_{\textrm{L}} < \Gamma_{\textrm{C}} \ll \Gamma_{\textrm{R}}$.

\section{Conclusion}
In conclusion, we have demonstrated the possibility to fabricate a tunable DQD in an InAs NW with highly sensitive and perfectly aligned charge readout QPCs by a single step wet etching process. The QPCs serving as charge detectors are strongly coupled to the DQD giving a remarkably large variation of typically $\sim 60~\%$ in the QPC conductance for changes of the charge states of the DQD. Simultaneous measurements using charge detection and measurement of the source-drain current through the DQD match exactly. We have presented time-resolved charge detection measurements which allows to track unambiguously individual electrons tunneling through the DQD in real-time and to determine the direction of the tunneling electrons. Both QPCs can detect all tunnel processes in the DQD exhibiting perfect correlation/anticorrelation of the signals. This enables us to distinguish clearly between QD-lead and interdot transitions. From the recorded time traces, all the tunnel rates of the involved tunnel processes can be extracted.

\addcontentsline{toc}{section}{Acknowledgements}
\section*{Acknowledgements}
We thank T. Ihn and A. Pfund for valuable discussions. Financial support from ETH Zurich is gratefully acknowledged.

\end{document}